\documentclass[preprint,12pt,3p]{elsarticle}

\usepackage{amssymb}
\usepackage{amsmath}

\usepackage{siunitx}
\usepackage{soul}

\journal{Tribology International}

\begin{document}

\begin{frontmatter}



\title{Anisotropic shrinkage and finite strains in confined frictional contacts}

\author[label1,label2]{Marco Ceglie}
\author[label1,label2]{Cosimo Mandriota}
\author[label1]{Giuseppe Carbone}
\author[label1]{Nicola Menga}   
\author[label2]{Antoine Chateauminois}
\affiliation[label1]{organization={Department of Mechanics, Mathematics and Management, Politecnico di Bari},
             addressline={Via Orabona, 4},
            city={Bari},
            postcode={70125},
          country={Italia}}
\affiliation[label2]{organization={Soft Matter Science and Engineering (SIMM), ESPCI Paris, PSL University, Sorbonne Université, CNRS UMR7615},
 addressline={10, rue Vauquelin},
city={Paris},
postcode={75005},
country={France}}
\begin{abstract}
We report on an experimental investigation of the interplay between friction, contact geometry and finite strains for smooth frictional contacts between rigid spherical glass probes and flat silicone substrates. Using both bulk and layered substrates under various loading conditions (normal force, radius of the probe), we show that shear-induced anisotropic shrinkage of the adhesive contact area under steady-state sliding is an effect of finite-elasticity conditions and is drastically affected by the level of geometric confinement. The resulting non-linear coupling between the normal and lateral directions is also investigated by measuring the changes in the indentation depth (conv. normal load) during the stiction of the adhesive contacts under imposed normal load (conv. indentation depth) conditions, with strong  effects of contact confinement. From a comparison with adhesiveless linear contact mechanics calculations, we show that the experimental observations can only be accounted for by the occurrence of finite strains/displacements conditions. Accordingly, measurements of the in-plane surface displacements at the surface of the rubber substrates confirm that strain levels well in the neo-Hookean range are experienced during steady-state frictional sliding.
\end{abstract}





\end{frontmatter}

%
\section{Introduction}
Friction of soft matter materials has long been recognized to induce strong changes in the shape of the contact area as compared to static adhesive equilibrium. Although these changes have many implications, especially regarding the determination of the actual contact area and the associated friction stresses in rough contact interfaces (see e.g.~\cite{Sahli2018,Weber2019}), their physical and mechanical origins still remains debated. When considering contacts between a rigid spherical probe and rubber substrates, a common observation is that shear loading during incipient stages of friction is resulting in an (anisotropic) reduction in the contact area~\cite{Barquins1993,Waters2010,Savkoor1977,Sahli2019} even in the absence of any significant viscoelastic effects or contact instabilities. Several contact mechanics models have been proposed so far to account for the shrinkage of adhesive contacts during the transition from rest to steady-state sliding, most of them being based on fracture mechanics arguments formulated within the framework of linear elasticity. Early studies by Savkoor and coworkers~\cite{Savkoor1977} focused on energy balance to predict a transition from JKR (Johnson, Kendall and Roberts) adhesive contacts~\cite{Johnson1971} to Hertzian contacts during stiction. However, this theory overestimates the reduction of the contact area. To better match the experimental results, Johnson and co-authors~\cite{Johnson1997,Kim1998} postulated an arbitrary interaction between friction and adhesion, so that the effective adhesion energy $\gamma_{\textrm{eff}}$ during sliding could be reduced and tuned to fit the experiments. Nonetheless, the contact shrinkage was still supposed to be axisymmetric, resulting in a circular area during sliding, as observed by Savkoor~\cite{Savkoor1977}. Only recently, averaging mode I,II, and III crack propagation modes (the so-called mixed mode) around the contact periphery, a model for the anisotropic contact area shrinkage has been proposed by Papangelo~\textit{et al.}~\cite{Papangelo2019}; however, it still requires to tune a phenomenological effective adhesion energy and mixed-mode functions (about six parameters) from a fit to experimental results.\\
Recently, geometrical and materials non-linearities emerged as an alternative to adhesion to account for the anisotropic changes in contact area under frictional sliding. Indeed, strain field measurements at the surface of silicon substrates in sliding contact with spherical glass probes showed that strain levels well beyond the linear range can be achieved~\cite{Nguyen2011,Chateauminois2017}.\textcolor{red}{ Theoretical studies by Lengiewicz~\textit{et al.}~\cite{Lengiewicz2020} and Wei~\textit{et al.}~\cite{Wei2024} have demonstrated that finite deformations alone can induce anisotropic shrinkage of soft contacts under shear, even in the absence of adhesion. In particular, finite element analyses of PDMS sphere–plane sliding show that linear elasticity predicts negligible change of contact area, while finite-strain hyperelastic models reproduce the observed anisotropy and trailing-edge detachment~\cite{Lengiewicz2020}. Moreover,the choice of the hyperelastic law (neo-Hookean, Mooney–Rivlin) alters the amplitude but not the qualitative features of contact area reduction, indicating that nonlinear elasticity is the governing mechanism. This quantitatively contrasts with earlier adhesion-masking models based on linear elasticity. In addition, these calculations especially showed that finite deformation can induce a significant coupling between the normal and lateral deformations modes which is not accounted for in small strain incompressible elasticity. Our own previous theoretical work has also shown that uniform tangential stresses do not change adhesion in linear theory, further ruling out adhesion as the primary origin of the shrinkage\cite{Menga2018,Menga2019b}.\\
In the present work, we do not attempt to develop a full finite-strain numerical model, but instead we aim to provide a systematic experimental comparison between bulk and confined PDMS substrates, thereby clarifying how finite-strain effects and the related coupling between normal and lateral loading are amplified by geometrical confinement.}
More precisely, we investigate the interplay between contact geometry and non-linear deformations of steady-state sliding contacts between smooth spherical probes and flat silicone substrates. For that purpose, the description of the contact shape under a wide range of contact conditions is combined with spatially resolved measurements of the associated surface strain field within the rubber. Moreover, we vary with the extent of non-linear effects also studying contacts involving silicone substrates of finite thickness. Additionally, the overlooked coupling between normal and lateral directions is investigated from sliding experiments carried out under either constant applied normal load or indentation depth conditions. As a baseline to evaluate the role of non-linearities in the observed contact responses, linear contact mechanics calculations are carried out which allows to determine the surface strain field and contact size for both semi-infinite and finite-sized contacts.\\
In the first section, we detail the experimental conditions and the contact-mechanics calculations. Then, observations of contact shape are detailed under imposed normal load conditions. These observations are subsequently extended to a comparison between imposed normal load and imposed vertical displacement conditions. In the last section, we discuss these observations in light of measurements of the surface strain fields.\\   
%
\section{Experimental details and calculations}
\subsection{Silicone rubber}
\label{sec:silicone_rubber}
\begin{figure} [!ht]
    \centering
    \includegraphics[width=0.8\linewidth]{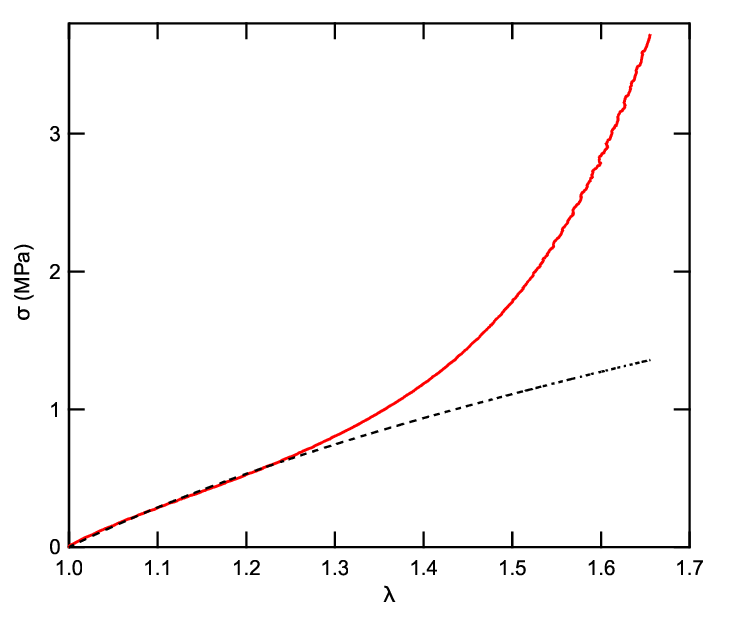}
   \caption{Nominal stress \textcolor{red}{$\sigma$} as a function of stretch ratio $\lambda$ for Sylgard 184 PDMS under uniaxial \textcolor{red}{tension} (crosshead velocity: 0.15~\si{\milli\meter\per\second}. The dotted line corresponds to a fit to neo-Hookean model for $\lambda \leq 0.3$ with $C_{10}$ = 0.526 \si{\mega\pascal}.}
  \label{fig:stress_strain_pdms}
\end{figure}

Experiments were carried out using PDMS rubber samples made of Sylgard 184 (Dow  Chemicals) mixed in a 10:1 weight ratio with its hardener and cured 48 hours at 70~\si{\celsius}. In addition to bulk specimens (with thickness $h>10$~\si{\milli\meter}, resulting in half-space contact behavior~\cite{Gacoin2006}), a S184 PDMS layer with a thickness $h=1.9$~\si{\milli\meter} was also manufactured to investigate the effects of mechanical confinement. After curing, this PDMS layer was glued on a rigid glass block. 
The Young's modulus $E$ and the adhesion energy $w$ of the PDMS material were determined from static indentation experiments on bulk specimen using JKR adhesive contact mechanics theory (see Supplementary Information). Consistently with previously reported results using the same Sylgard 184 system~\cite{Varner2024}, the measured Young's modulus and adhesion energies were observed to vary slightly from one specimen to another ($E=2.5 \pm 0.6$~\si{\mega\pascal}, $w=30 \pm 15$~\si{\milli\joule\per\square\meter}). Additional indentation experiments were carried out with the layered substrates (\textcolor{red}{see Supplementary Information}). No significant difference in the elastic properties of the PDMS layer was observed when these indentation data were analyzed using a previously developed adhesive contact model for coated substrates~\cite{Mary2006}.\\
For the purpose of measuring the in-plane surface displacement fields, specimen were surface-marked with a square network of cylindrical holes 10~\si{\micro\meter} in diameter and a mesh size of 70~\si{\micro\meter} using conventional lithography tools. These holes were acting as markers which were tracked during the sliding friction experiments using a previously developed methodology~\cite{Nguyen2011}.\\
Uniaxial extension tests were performed at a displacement rate of 0.15~\si{\milli\meter\per\second} to characterize the non-linear behavior of the silicon rubber up to failure. The nominal stress $\sigma$ vs. stretch $\lambda$ behavior is shown in Fig.~\ref{fig:stress_strain_pdms}, together with a fit of the experimental data to a neo-Hookean model for $\lambda <1.25$ (dotted line):
\begin{equation}
   \sigma= 2 C_{10} \left(\lambda - \frac{1}{\lambda^2}\right),
\end{equation}
with $C_{10}=0.526$~\si{\mega\pascal}. Results clearly show that material response significantly deviate from linearity, and that simple hyperelastic models (e.g., neo-Hookean), though non-linear, fail to predict the real material behavior for $\lambda \gtrsim 1.3$.
%
\subsection{Friction experiments}
\label{sec:friction_experiments}
\begin{figure} [!ht]
    \centering
    \includegraphics[width=1\linewidth]{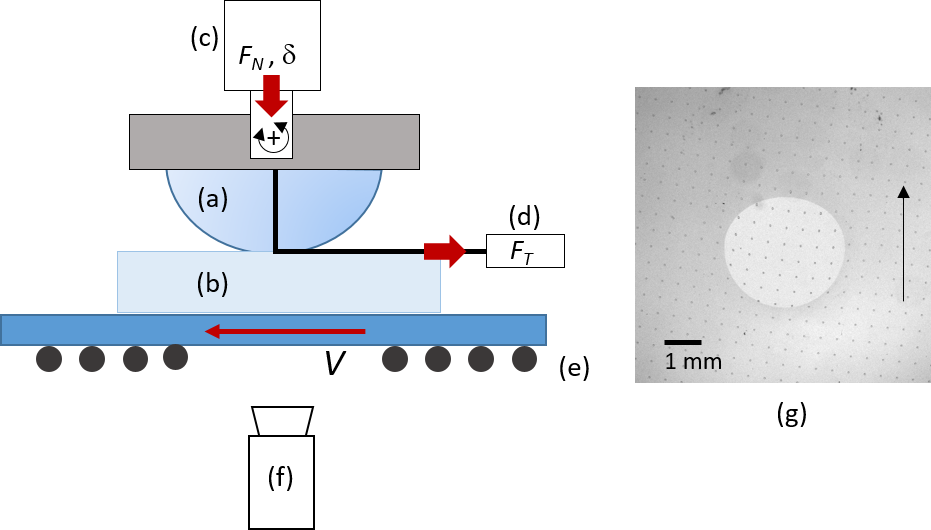}
    \caption{Schematic description of the friction set-up. A smooth glass lens (a) is contacting a smooth, surface marked, PDMS substrate (b). An electromagnetic actuator (c) allows to impose either normal load ($F_N$) or indentation depth ($\delta$) conditions during sliding. The PDMS substrate is displaced with respect to the fixed glass lens at imposed velocity $V$ by means of a motorized translation stage (e). The normal load $F_T$ is recorded within the contact plane by means of a strain gage load sensor (d). A zoom lens and a CCD camera (f) allows to record contact images (g).}
    \label{fig:setup}
\end{figure}

All friction experiments were carried out using smooth optical glass lenses with radii of curvature $R=$~9.3 and 20.7~\si{\milli\meter} and a home-made device schematically described in Fig.~\ref{fig:setup}. 
%
An electromagnetic actuator operated in a feedback loop allowed to perform sliding experiments under either imposed normal load ($F_N=$~0.02 to 8~\si{\newton}) or imposed indentation depth conditions. For practical reasons, sliding experiments under imposed indentation depth were restricted to the stiction phase, \textit{i.e.} to the transition from static adhesive contact to gross sliding. Indeed, it was experimentally very difficult to ensure that the surface of the PDMS specimen is perfectly (\textit{i}) flat and (\textit{ii}) aligned with respect to the translation axis of the stage. As a consequence of these small defects, it was not possible to ensure a constant indentation depth when the lateral displacement exceeded the contact size. However, such a lateral displacement was found enough to induce gross sliding at the contact interface without any significant change in the actual indentation depth.\\ 
In addition to the continuous monitoring of the lateral, $F_t$, and normal, $F_n$, forces, contact visualization was performed through the transparent PDMS substrate by means of a zoom lens and a CMOS camera (2048x2048 pixels, 8 bit resolution) operated in a light transmission mode, allowing to measure the contact area in both the static ($A_0$) and sliding ($A_S$) conditions together with the surface displacement field. Particle Image Velocimetry (PIV) tracking of the dot markers allowed to map the displacements $u$ and $v$ along and perpendicular to the sliding direction, respectively. The resulting space resolution for the displacement fields is about 10~\si{\micro\meter}.\\

\begin{figure} [htbp]
   \centering
    \includegraphics[width=0.95\linewidth]{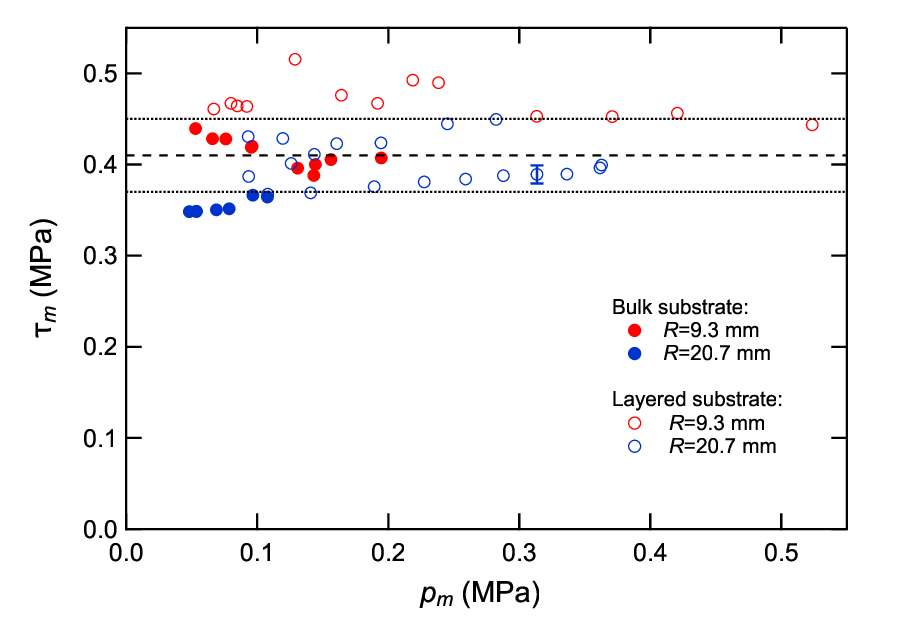}      
   \caption{\textcolor{red}{Average steady-state frictional shear stress $\tau_m=F_t/A_s$ as a function of the mean contact pressure $p_m=F_N/A_s$ where $A_s$ is the steady-state sliding contact area for bulk and layered substrates (Radius of the glass lens: $R=9.3$~\si{\milli\meter} and $R=20.7$~\si{\milli\meter}).} Large and small dotted horizontal lines correspond to the average value of $\tau_m$ and its standard deviation, respectively. Scatter mostly originates from differences between different samples.}
 \label{fig:tau_pm}
\end{figure}

Keeping in mind that the frictional stress of the used PDMS material was observed to be only weakly (logarithmically) increasing with velocity~\cite{Fazio2021}, we decided to carry out experiments at a single imposed velocity, $v=0.1$~\si{\milli\meter\per\second}. For the selected contact conditions, the measured contact radius $a$ is of the order of one millimeter, so that the characteristic frequency involved in bulk material deformation during sliding can be estimated to be $\omega_b = v/a \lessapprox 0.1$~\si{\per\second}, \textit{i.e.} about nine orders of magnitude less than the frequency of the glass transition of the used Sylgard 184 silicone at room temperature ($10^8$~\si{\hertz})~\cite{Nguyen2013}. As a result, any substantial viscoelastic effects can be discarded from the observed changes of the contact shape during sliding. The temperature of the laboratory environment was $24 \pm 2$~\si{\celsius} and the relative humidity ranged between 35\% and 45\%.\\
Figure~\ref{fig:tau_pm} shows the measured average frictional shear stress $\tau_m=F_t/A_s$ during steady state sliding as a function of the mean contact pressure $p_{m}=F_n/A_s$ for the investigated lens radii and specimen thicknesses, where $F_t$, $F_n$ and $A_s$ denote the steady-state frictional force, normal load and contact area, respectively. Here, the scatter mostly originates in slight differences which are invariably observed between the frictional properties of different specimens. Consistently with previous findings~\cite{Chateauminois2008,Gao2004}, $\tau_m$ is observed to be independent on contact pressure within experimental accuracy, with no measurable effects of the radius of the lens and of the substrate's thickness. The average frictional shear stress is $\tau=0.41 \pm 0.04$~\si{\mega\pascal}.\\
%
\subsection{Contact mechanics calculations}
\label{sec:contact_mechanics_calculations}
Linear elastic contact mechanics calculations were performed as a benchmark to predict the contact area and the displacement field during frictional sliding. Steady-state gross slip conditions are assumed. Since most of the existing linear models for sliding adhesive contacts postulate that friction (partially) masks adhesion, leading to a reduction in the effective energy of adhesion $\gamma_{\textrm{eff}}$ \cite{Savkoor1977,Johnson1997,Kim1998,Papangelo2019}, as a worst case scenario, to assess linear predictions we assumed adhesiveless frictional sliding (\textit{i.e.}, $\gamma_{\textrm{eff}}=0$).
Previous investigations by Chateauminois~\textit{et al.}~\cite{Chateauminois2008,Chateauminois2017} showed that the frictional shear stress distribution within smooth contact interfaces between PDMS and rigid probes (\textit{i}) is pressure independent and (\textit{ii}) depends on the in-plane surface stretch of the rubber. Here, we neglected this stretch effect and assumed an uniform shear stress distribution within the contact, so that the interfacial shear stress along the sliding direction is $\tau(\textbf{r})=\tau_m$ at any given point $\textbf{r}$ in the contact interface.\\
For thick, incompressible, substrates, Green's tensor~\cite{landau1986} shows that no interaction between in-plane and out-of-plane elastic fields are involved within the framework of linear elasticity. The value of the sliding contact area is thus simply given by the Hertz theory, while analytical solutions for the in-plane displacements fields can be found in Ref.~\cite{Menga2019}.\\
Conversely, when dealing with layered substrates a coupling exists between in-plane and out-of-plane elastic fields, even within the framework of linear elasticity. Therefore, frictional sliding contacts differ from static ones, as recently investigated in a series of studies~\cite{Menga2019a,Menga2020,Menga2021,Muller2023}. In this situation, both the frictional contact area and the in-plane displacements fields are calculated using the 3D extension of the Green's Function Molecular Dynamics (GFMD) method provided in Ref.~\cite{Muller2023}.\\
%
\section{Results and discussion}
\subsection{Contact shape under steady-state sliding}
\begin{figure} [htbp]
   \centering
    \includegraphics[width=1\linewidth]{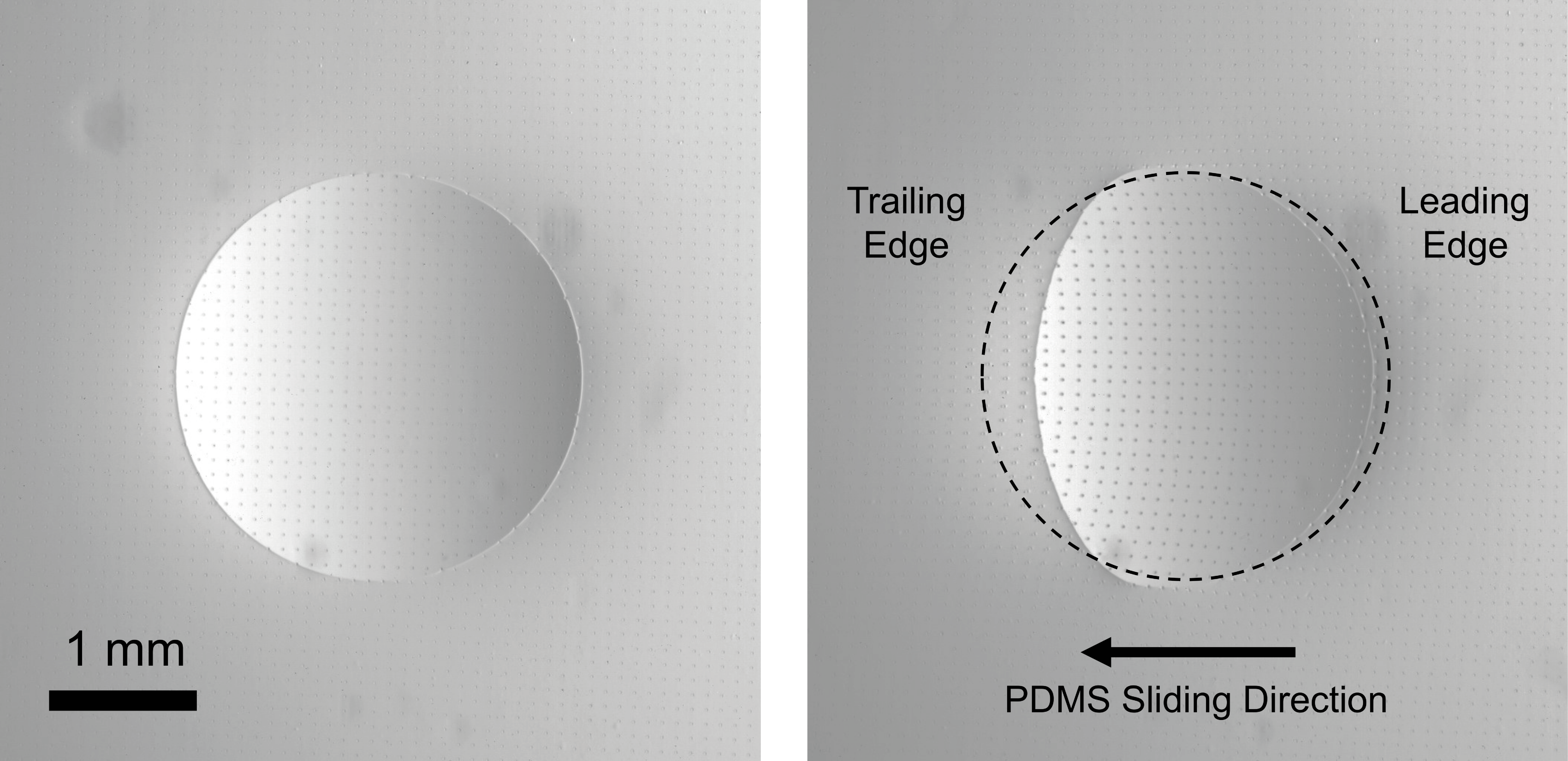}       
   \caption{Left: Static adhesive contact picture under normal indentation. Right: steady-state sliding contact picture, superimposed to the static contact boundary as a dashed line (Bulk S184, $R=20.7$~\si{\milli\meter}, $F_N=$~0.5~\si{\newton}). The silicon substrate is moving from right to left with respect to the fixed glass lens.}
 \label{fig:contact_pictures}
\end{figure}

\begin{figure*} [htbp]
  \includegraphics[width=1\textwidth]{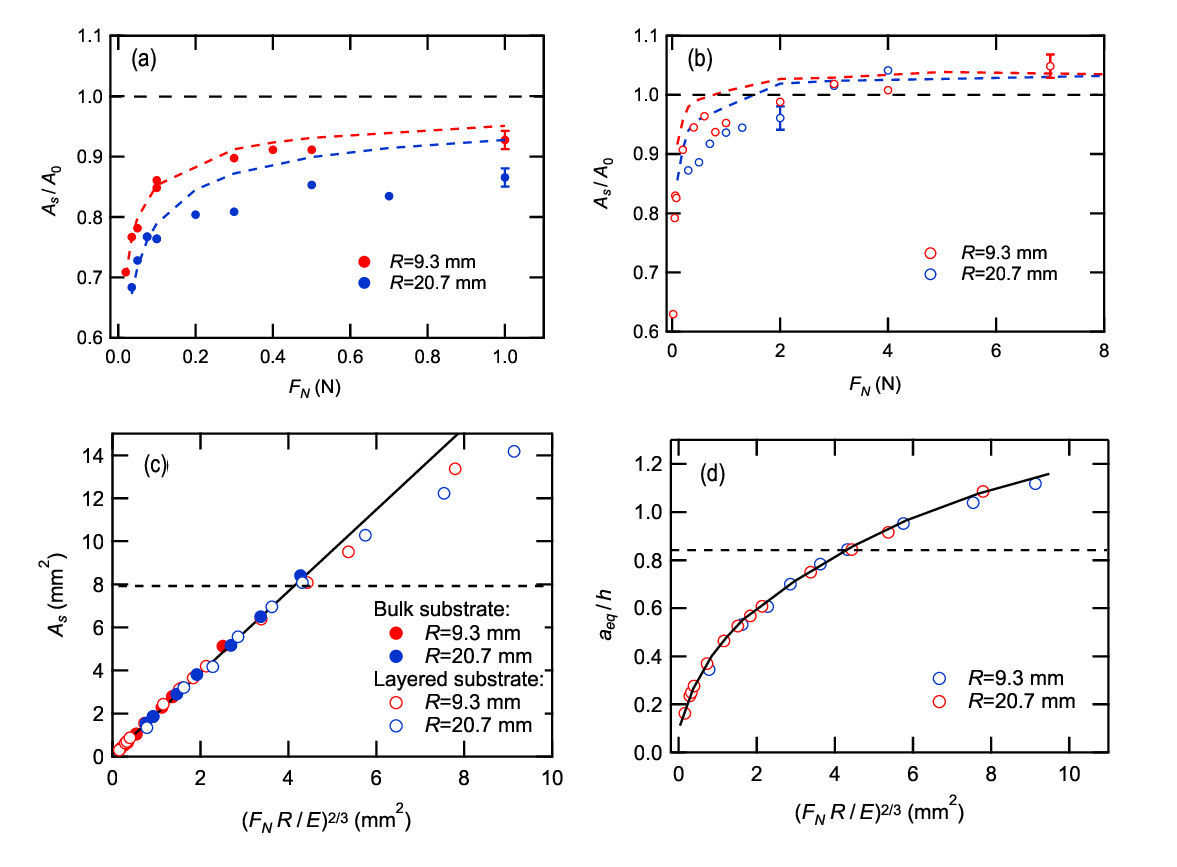}
  \caption{\textcolor{red}{Normalized sliding contact area $A_s/A_0$ (where $A_0$ is the static contact area) as a function of the normal load $F_N$ for (a) bulk and (b) layered substrates (lens radius $R=9.3$~\si{\milli\meter} and $R=20.7$~\si{\milli\meter}. Dotted lines correspond to theoretical adhesiveless contact mechanics predictions within the framework of linear elasticity (see text for details). (c) measured sliding contact area $A_s$ as function of $(F_{N}R/E)^{2/3}$ for both the bulk and layered specimens for $R=9.3$~\si{\milli\meter} and $R=20.7$~\si{\milli\meter}}. The solid line is a linear regression based on bulk substrate data. (d) normalized equivalent contact radius $a_{\textrm{eq}}/h$ during steady state sliding as function of $(F_{N}R/E)^{2/3}$ for the layered specimen measurements, with $a_{\textrm{eq}}=\sqrt{A_{s}/\pi}$ and $h$ the layer thickness. The solid line is a guide for the eye. Mechanical confinement effects for the layered specimen are significative for $a_{eq}/h\gtrsim0.85$, \textit{i.e.}, above the horizontal dashed line in (c) and (d).}
  \label{fig:area_sliding}
\end{figure*}

In this section, we report on the size and shape of steady-state sliding contacts under imposed normal load conditions. As a typical example, Figure~\ref{fig:contact_pictures} shows images of a contact in static adhesive equilibrium (left) and during steady-state sliding (right, with the flat PDMS substrate moving from right to left with respect to the fixed glass lens). In the right image, the edge of the static adhesive contact has also been superimposed as a dashed circle to the image of the sliding contact. Consistently with previous observations for similar contacts between spherical glass probes and rubber substrates~\cite{Barquins1993,Waters2010,Savkoor1977,Sahli2019}, it turns out that the contact area undergoes an anisotropic shrinkage with respect to the static adhesive equilibrium. Although some loss of contact is observed at the leading edge of the contact, the predominant effect is a detachment of the silicone from the glass probe at the trailing edge.\\
%
Figures~\ref{fig:area_sliding}(a,b) report the measured steady-state contact area $A_s$ normalized with respect to the measured static contact area $A_0$ for different lens radii as a function of the applied normal force $F_N$, for bulk and layered PDMS specimens, respectively. In addition to experimental data, the theoretical predictions for adhesiveless, linear elastic, frictional contacts are also reported as dotted lines. As detailed above (section~\ref{sec:contact_mechanics_calculations}), the prediction for $A_s$ in the case of bulk specimens (dotted lines in Figure~\ref{fig:area_sliding}(a)) simply corresponds to Hertzian calculation as there is no coupling between normal and lateral directions (here, theoretical $A_s$ values are normalized with respect to the static adhesive values $A_0$ deduced from JKR theory). Conversely, for layered substrates, in-plane and out-of-plane coupling occurs even in the framework of linear contact mechanics theories. This coupling is accounted for in our GFMD BEM model~\cite{Muller2023} which was used to predict $A_s$ (the corresponding $A_s$ values were normalized with respect to the prediction $A_0$ of our static adhesive contact model for layered substrates~\cite{Mary2006}).\\ 
%
As shown in Figure~\ref{fig:area_sliding}(a), for the thick incompressible PDMS substrate, the steady-state sliding contact area $A_s$ is lower than the static adhesive contact area for all the considered normal load range. Moreover, measured $A_s$ values are significantly lower than the adhesiveless frictional value given by the Hertzian theory, except maybe at the lowest loads, $F_N<0.1$~\si{\newton}. This clearly indicates that a reduction in adhesive contributions is not an adequate argument to explain the decrease in contact area: the case at hand can no longer be discussed assuming linear elasticity, as pointed out in \cite{Menga2018MCD,Menga2019MCDcorr} and, eventually, shown in~\cite{Lengiewicz2020}.\\
Figure~\ref{fig:area_sliding}(b) focuses on the case of layered substrates. As opposed to bulk substrates, experimental $A_s$ values can exceed that corresponding to the adhesive equilibrium $A_0$ when $F_N$ is increased. Such an effect is consistent with theoretical contact mechanics calculations in Refs.~\cite{Menga2019a,Menga2021,Muller2023} which showed that confinement induced coupling reduces the contact stiffness as compared to static frictionless contacts and that the effect of adhesiveless frictional sliding can downright exceed static adhesion, so that $A_s/A_0>1$. Nonetheless, the overall physical picture for measured sliding conditions is qualitatively similar to what observed for bulk substrates: regardless of the normal force, the measured contact area during sliding $A_s$ is smaller than the adhesiveless linear theory prediction, indicating that non-linear effect cannot be neglected.\\
Though $A_s$ is different from the adhesiveless elastic calculation, Figure~\ref{fig:area_sliding}(c) shows that, regardless of the substrate thickness, $A_s\propto(F_NR/E)^{2/3}$ up to a certain threshold ($(F_NR/E)^{2/3}\approx4$). This means that below this threshold Hertzian dimensional arguments still qualitatively hold true during sliding, \textit{i.e.} even beyond the limit of linear elasticity. For $(F_NR/E)^{2/3}\gtrapprox4$, the layered substrate behavior diverts from Hertzian scaling. The reason for this can be inferred from Figure~\ref{fig:area_sliding}(d), showing the ratio between the equivalent contact radius $a_{\textrm{eq}}=\sqrt{A_s/\pi}$ and the substrate's thickness $h$ as a function of $(F_NR/E)^{2/3}$. Indeed, the threshold $(F_NR/E)^{2/3}\approx4$ corresponds to $a_{eq}/h\approx 1$ which is the critical condition for confinement to significantly affect the contact response, limiting the possibility of the layered incompressible solids to accommodate the displacements remotely along its thickness.\\

\begin{figure*} [htbp]
  \includegraphics[width=1.0\textwidth]{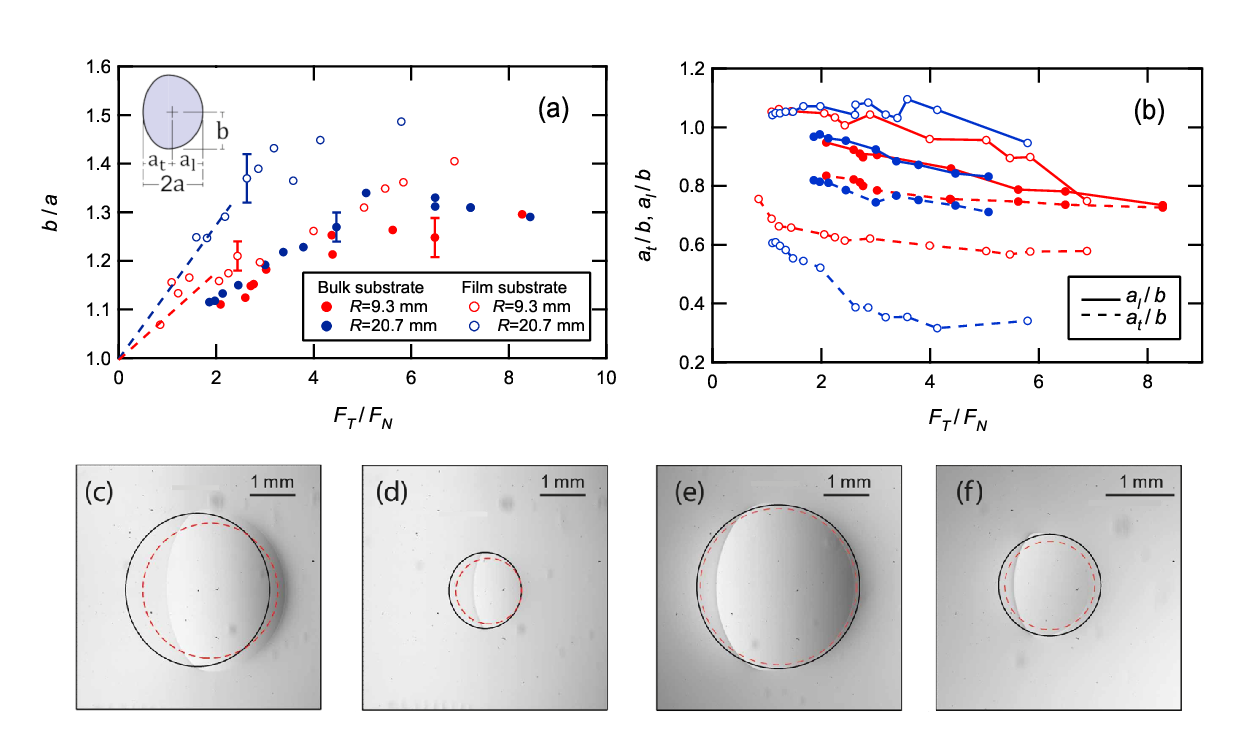}
\caption{Anisotropy of the contact area under steady-state sliding for bulk and layered substrates. (a) Contact aspect ratio $b/a$ versus normalized friction force $F_T/F_{N}$ for $R=9.3$~\si{\milli\meter} and $R=20.7$~\si{\milli\meter}. $F_T/F_{N}$ is approximately proportional to $F_{N}^{-1/3}$ (see the text for further detail). The dotted lines correspond to linear fits for $F_T/F_{N} \leq 0.2$ with the intercept set to unity for layered substrates. (b) Normalized distances of the contact trailing edge $a_{t}/b$ and leading edge $a_{l}/b$ from the center of the spherical indenter as a function of the the normalized friction force $F_T/F_{N}$. All the geometrical quantities are defined in the inset in (a). (c)-(f) averaged images of the contact area under steady state sliding for layered (c,d) and bulk (e,f) specimens for $F_{N}=1$~\si{\newton} (c,e) and $F_{N}=0.1$~\si{\newton} (d,f). Black lines delimit the static contact area; dashed red lines correspond to linear elastic predictions of the contact shape under steady state sliding.}
 \label{fig:contact_anisotropy}
\end{figure*}

Aside from the shrinkage of the contact area investigated in Figure~\ref{fig:area_sliding}, the non-linearity induced by frictional sliding also affects the contact shape in terms of anisotropy. As shown in the inset of Figure~\ref{fig:contact_anisotropy}(a), we define $2a$ and $2b$ as the longitudinal and transverse dimension of the sliding contact area, respectively, while $a_{t}$ and $a_{l}$ are the distance of the contact trailing and leading edges, respectively, with respect to the center of the spherical lens (undeformed static contact area).
%
Figure~\ref{fig:contact_anisotropy}(a) shows the contact aspect ratio $b/a$ as a function of the measured normalized friction $F_T/F_N$. According to Figure~\ref{fig:tau_pm} and Figure~\ref{fig:area_sliding}(c), we have the following scaling for $F_T/F_N = \tau_m A_s/F_N \propto F_N^{-1/3}$. Focusing on the bulk case, we recall that linear theory predicts vanishing coupling between normal and tangential loads and displacements; therefore, $b/a$ should be unity value regardless of $F_T/F_N$. In the layered case, elastic contact mechanics calculations indicate that coupling is non-vanishing, though theoretical calculations show that the contact shape anisotropy remains negligible (about $1\%$). To set the stage, the behavior reported in Figures~\ref{fig:area_sliding}(a,b) and \ref{fig:contact_anisotropy}(a) basically indicates that contact shrinkage and anisotropy cannot be accounted for within the framework of linear elasticity.\\
More specifically, we report generally higher contact anisotropy (\textit{i.e.}, $b/a>1$) for thin layers compared to the bulk case. Moreover, in the layered case, the larger lens radius results in a larger contact size and, in turn, in larger confinement effects, which trigger a higher contact anisotropy, compared to the lower radius which, for relatively low values of $F_N$, behaves similarly to the bulk case. More importantly, for the layered case we expect that $b/a$ should approach asymptotically unity at large values of $F_N$ (as schematically indicated by dotted lines in Figure~\ref{fig:contact_anisotropy}(a)), \textit{i.e.} for highly confined contact conditions where contact shape should depend on Hertzian response of the glass substrates.\\
In Figure~\ref{fig:contact_anisotropy}(b) we detail the friction-induced shift of the contact leading and trailing edges with respect to the center of the spherical glass probe. Overall, the contact area shifts in the direction of sliding (see Figures~\ref{fig:contact_anisotropy}(c-f). However, regardless of the confinements, this behavior is mostly related to the shift of the trailing edge along the sliding direction, as $a_{t}/b$ is systematically lower than $a_{l}/b$. Confinement seems to enhance this shift of the contact area, which is also predicted by linear contact mechanics calculations (see red dotted lines in Figures~\ref{fig:contact_anisotropy}(c,d)) in the absence of any significant contact anisotropy.\\
Though anisotropy is not expected in linear theory, qualitative arguments (see \cite{Muller2023}) clearly indicate that, even in linear theory, in-plane surface stresses are compressive close to the leading edge and tensile at the trailing edge, regardless of the confinement. In-plane tensile stress and finite displacements at the trailing edge of the contact are likely to induce a downward displacements of the free surface as a result of the reduced surface curvature, thus shifting the contact trailing edge towards the spherical lens center (\textit{i.e. }reducing $a_{t}/b$). In addition, Poisson's contraction of the stretched incompressible rubber can also induce downward displacements at the trailing edge.\\
Interestingly, in the layered specimen case, the non-linear coupling effect also leads to $a_{l}/b>1$ when confinement is increased (\textit{i.e.} at low $F_T/F_N \propto F_N^{-1/3}$ values). In other words, the leading edge's boundary has moved from left to right once the steady-state sliding condition has been established (see Fig. \ref{fig:contact_anisotropy} (e)). This observation can be related to the compressive state of the confined material close to the leading edge which could induce an out-of-plane bulge of material. Similar trends have been recently reported in Ref. \cite{Wei2024} while, in a different context, the buckling phenomenon at the leading edge has been invoked to account for the generation of Schallamach waves in rubber contacts~\cite{WuBavouzet2007,Viswanathan2015}. Here, it could be enhanced by the geometrical confinement of the incompressible silicone layer between the rigid glass substrates. Accordingly, axisymmetric material pile-up due to incompressibility close to the contact edges is reported even in purely normal contacts for nearly incompressible layers lying on rigid substrates~\cite{Barthel2006}; therefore, at very large normal load, the symmetric effect is dominant so that the leading and trailing edges shift in along the sliding direction almost by the same quantity (\textit{i.e.}, $a_l+a_t \approx 2b$). Nonetheless, although increasing the applied normal force has a stabilizing effect on the aspect ratio $b/a$ (asymptotically tending to unit value - see Figure~\ref{fig:contact_anisotropy}(a)), the contact shape is always asymmetric and never recovers the circular shape (see Figs. \ref{fig:contact_anisotropy} (c)-(f)).
\subsection{Imposed normal load vs. displacement conditions}
\begin{figure*} [htbp]
  \includegraphics[width= 1.0 \textwidth]{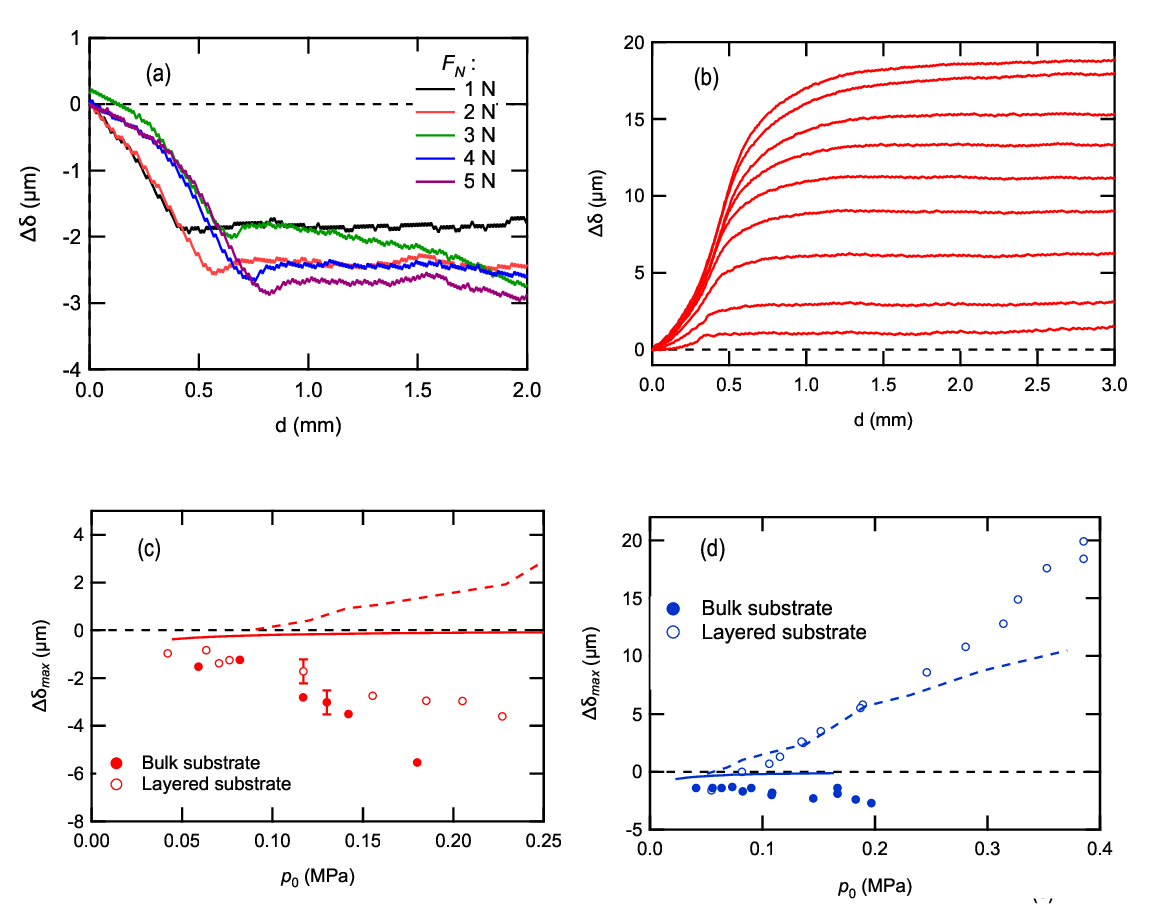}
   \caption{Stiction under imposed normal load conditions. \textcolor{red}{Change in the indentation depth $\Delta \delta$ as function of the sliding distance $d$ for (a) bulk and (b) layered substrates (lens radius $R=20.7$~\si{\milli\meter}). In (a) the normal load $F_N$ is increased from 1 to 5~\si{\newton} by 1~\si{\newton} increments. In (b), the applied normal load $F_N$ is increased from 1~\si{\newton} to 8~\si{\newton} by 1~\si{\newton} increment from bottom to top. The magnitude $\Delta \delta_{max}$ of the change in the indentation depth during stiction for bulk and layered substrates is shown as a function of the nominal static contact pressure $p_0$ for (c) $R=9.3$~\si{\milli\meter} and (d) $R=9.3$~\si{\milli\meter}}. Colored continuous and dotted lines correspond to linear theoretical prediction for bulk and layered substrates, respectively.}
   \label{fig:delta_z}
\end{figure*}

As suggested by the observed anisotropic shrinkage of the contact area, non-linear finite strains effects activated during frictional sliding lead to a certain degree of in-plane and out-of-plane coupling even for nominally uncoupled systems, such as bulk incompressible samples (with vanishing second Dundurs' constant). In this section, we investigate how this coupling is affected by the imposed contact conditions in the normal direction, namely either imposed normal force or normal indentation depth. For practical reasons detailed in section~\ref{sec:friction_experiments}, this analysis was restricted to the stiction phase, \textit{i.e.} to the transition from adhesive static equilibrium to gross slip conditions.\\
Figure~\ref{fig:delta_z}(a), shows the changes in the indentation depth $\Delta \delta$ as a function of the imposed lateral displacement $d$, for a bulk PDMS substrate in contact with a lens of radius $R=20.7$~\si{\milli\meter} under imposed normal load conditions. Here, positive $\Delta \delta$ values corresponds to an upward displacement of the glass lens. Accordingly, the spherical indenter is lifted-up during stiction of bulk substrates with a limited effect of the applied normal load. 
%
Conversely, stiction is found to induce an increase in the indentation depth during stiction of the layered substrate with a marked effect of the applied normal force (Figure~\ref{fig:delta_z}(b)).\\
In Figures~\ref{fig:delta_z}(c,d) the changes $\Delta \delta$ in the indentation depth at the onset of full sliding are reported as a function of nominal contact pressure $p_0=F_N/A_0$ for both bulk and layered substrates and for the two different lens radii. In addition, theoretical prediction within the framework of linear contact mechanics are shown as colored continuous and dotted lines for the bulk and layered substrates, respectively. For bulk substrate, the change in the theoretical indentation depth is assumed to arise only from the JKR to Hertz transition while for the layered substrates it accounts for the linear coupling between normal and lateral loads as a result of contact confinements. 
For the indenter with $R=9.3$~\si{\milli\meter}, a lifting-up of the indenter is observed for both the bulk and layered substrates (Figure~\ref{fig:delta_z}(c)). However, a small reduction in the absolute value of $\Delta \delta_{max}$ is observed for the layered specimen as compared to the bulk substrate when the contact pressure, \textit{i.e.} the contact confinement is increased. For the bulk substrate, the theoretical prediction shows that the observed change in the indentation depth can clearly not be ascribed to the JKR to Hertz transition.\\
For the indenter  $R=20.3$~\si{\milli\meter}, at low contact pressure, $a_\textrm{eq}<h$ and the contact behavior of layered specimens is similar to that of bulk samples (Figure~\ref{fig:delta_z}(d)). Conversely, as compared to the lens with $R=9.3$ ~\si{\milli\meter}, much higher confinements are achieved in the layered specimen when the contact is pressure is increased with $R=20.7$~\si{\milli\meter} which induce an increasing sinking-in of the glass lens which is semi-quantitatively consistent with the theoretical calculation.\\

\begin{figure} [!ht]
	   \centering
  \includegraphics[width= 0.8 \textwidth]{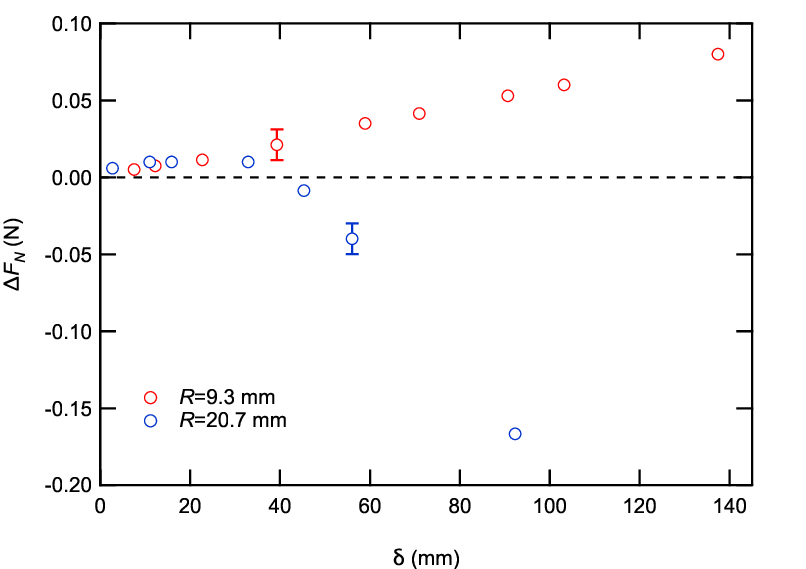}
\caption{Magnitude of the change in the normal compressive force $\Delta F_{N}$ (positive if increasing) as a function of the imposed indentation depth $\delta$ during stiction of a layered PDMS substrate for $R=9.3$~\si{\milli\meter} and $R=20.7$~\si{\milli\meter}.}
     \label{fig:delta_fn}
\end{figure}

A specular scenario is depicted in Fig.\ref{fig:delta_fn}, which reports change in the normal force $\Delta F$ during stiction under fixed normal penetration $\delta$, for layered samples. In this case, when the applied pressure is increased, the effective contact stiffness reduces for $R=20.7$~\si{\milli\meter} due to coupling in confined situations, and slightly increases for $R=9.3$~\si{\milli\meter} (basically behaving like the bulk case).\\
%

\textcolor{red}{At this stage, some connection to finite-strain elasticity can be done by considering a qualitative benchmark against the FE results by Lengiewicz~\textit{et al}~\cite{Lengiewicz2020} for the sliding of a smooth, neo-hookean, hemisphere supported by a pillar on a smooth glass substrate/glass under a constant, pressure-independent Tresca-type interfacial shear stress. Specifically, it turns out that our experimental observations for a pressure-independent shear strength ($\tau_m \approx$~0.41~\si{\mega\pascal}) - namely ellipse-like anisotropy, dominant trailing-edge detachment and $\delta \Delta $ sign change with confinement- are consistent with the mechanisms identified by that model (finite-deformation-induced lifting at the trailing edge, negligible area change under linear elasticity, and strong normal–tangential coupling).}

\subsection{Displacement and strain fields}
\begin{figure} [!ht]
   \centering
    \includegraphics[width=1\linewidth]{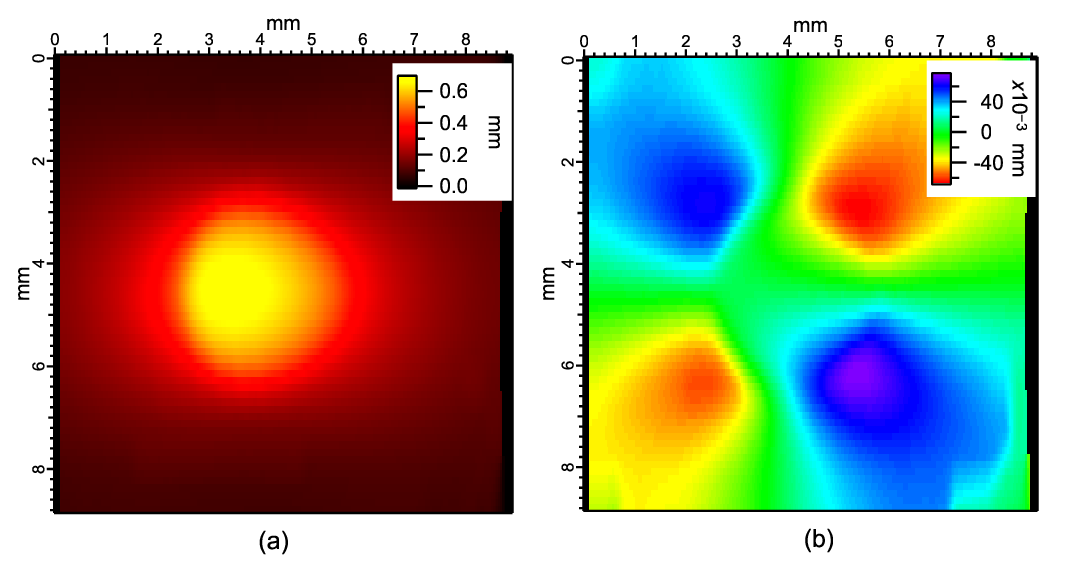}
   \caption{Surface displacement field of the bulk silicone substrate during steady-state sliding ($F_N=1$~\si{\newton}, $R=20.7$~\si{\milli\meter}). (a) displacement component $u_x$ along the sliding direction, (b) displacement component $u_y$ perpendicular to the sliding direction. The rubber substrate is moved from right to left with respect to the fixed spherical probe, and space coordinates refer to the undeformed condition.}
 \label{fig:displacement_field_bulk}
\end{figure}

\begin{figure} [!ht]
   \centering
   \includegraphics[width=0.7\linewidth]{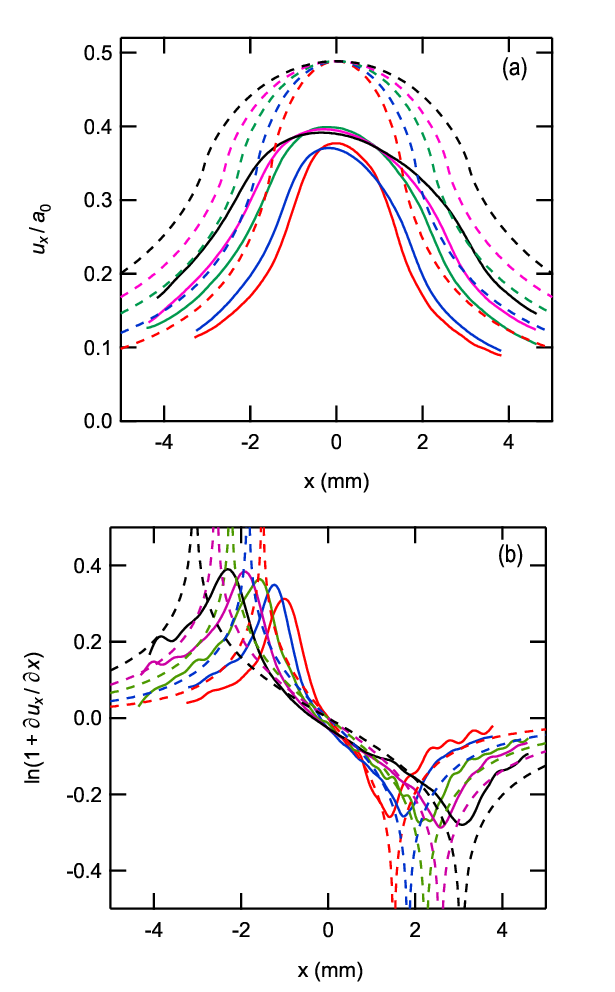}
   \caption{Profiles of (a) dimensionless surface displacement $u_x/a_0$ \textcolor{red}{(where $u_x$ is the longitudinal surface displacement and $a_0$ the static contact radius)} and (b) of in-plane-logarithmic strain taken across the contact and along the sliding direction for different normal loads $F_N$ for the thick substrate (lens radius $R=20.7$~\si{\milli\meter}). $F_N=0.5$~\si{\newton} (red), 1~\si{\newton} (blue), 2~\si{\newton} (green), 3~\si{\newton} (purple) and 5~\si{\newton} (black). Profiles are mapped in Eulerian coordinates, \textit{i.e.} relative to the equilibrium state after the application of lateral displacement. Continuous and dashed lines are for measurements and linear predictions, respectively.}
 \label{fig:strain_profiles_bulk}
\end{figure}

\begin{figure} [!ht]
   \centering
    \includegraphics[width=0.7\linewidth]{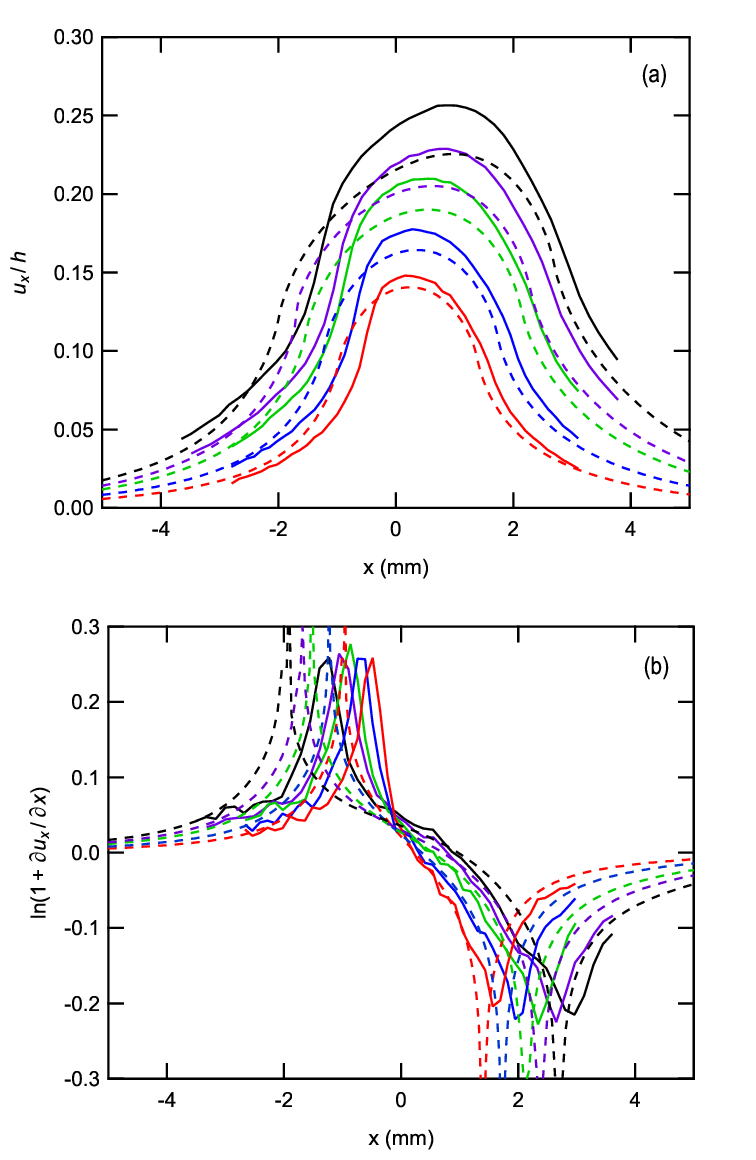}    
   \caption{Profiles of (a) dimensionless in-plane surface displacement $u_x/h$ \textcolor{red}{(where $u_x$ is the longitudinal surface displacement and $a_0$ the static contact radius)} and (b) of in-plane logarithmic strain taken across the contact and along the sliding direction for different normal load s$F_N$ for a confined contact with a S~184 layer of thickness $h=1.9$~\si{\milli\meter} (lens radius $R=20.7$~\si{\milli\meter}). $F_N=0.5$~\si{\newton} (red), 1~\si{\newton} (blue), 2~\si{\newton} (green), 3~\si{\newton} (purple) and 5~\si{\newton} (black). Profiles are mapped in Eulerian coordinates, \textit{i.e.} relative to the equilibrium state after the application of lateral displacement. Continuous and dashed lines are for measurements and linear predictions, respectively.}
 \label{fig:strain_profiles_thin}
\end{figure}

In this section, we investigate the in-plane displacement and strain fields once steady-state sliding conditions are fully achieved with a pressure independent shear stress. The measured fields at the polymer's surface will be compared to the linear predictions. The latter are derived based on analytical solution for the thick (half-space) case \cite{Menga2019}, while numerical calculations using the aforementioned GFMD method \cite{Muller2023} have been performed for layered substrates. In Fig. \ref{fig:displacement_field_bulk}, the measured in-plane displacement fields are shown for the bulk substrate subjected to a normal force $F_N=1$~\si{\newton}. As expected, due to frictional sliding, the substrate mostly displaces in the sliding direction (i.e., along the $x$ axis) (Fig. \ref{fig:displacement_field_bulk}a), while the transverse displacement displays a quadripolar symmetry \cite{Menga2019} as a result of (non-linear) Poisson's effects (Fig. \ref{fig:displacement_field_bulk}b) (see also~\cite{Nguyen2011}). 
%
Profiles of the measured and calculated normalized longitudinal displacement $u_x/a_0$ (with $a_0$ the radius of the static adhesive contact) and of the in-plane logarithmic strain $\ln(1+\partial u_x/\partial x)$ taken along the sliding direction $x$ at the contact mid-section are also shown in Fig.~\ref{fig:strain_profiles_bulk}. Since linear theory predicts $u_x \propto a_0$ for bulk samples \cite{Menga2019}, we show the dimensionless quantity $u_x/a_0$ as an estimate of the average shear strain $\epsilon_{xz}$ in the contact region.\\ 
%
However, though linear theory predicts a symmetric behavior with respect to the contact area center, measured displacements clearly show a significant asymmetry, with the maximum normalized displacement $u_x/a_0$ being shifted towards the contact trailing edge. As a consequence, the in-plane tensile elastic strain at the trailing edge is significantly larger than the compressive one at the leading edge, as shown in Fig.~\ref{fig:strain_profiles_bulk}(b). Of course, the strain singularity at the contact edge predicted in linear theory depends on the step-change in the shear stress at the contact boundary, while the real distribution is smoother, though very steep. The absolute values of the maximum measured maximum in-plane surface tensile/compressive strain are in the range $0.3\pm 0.05$, \textit{i.e.} well in the non-linear regime of the silicone elastomer and even slightly above the neo-Hookean regime (see Fig.~\ref{fig:stress_strain_pdms}). Accordingly, the observation that the measured shear and tensile strains are lower than the calculated ones can be accounted for by the sub-linear, neo-Hookean, behavior of the PDMS in the considered strain range. Noticeably, the measured strains are only weakly dependent on the applied normal load.\\
Figure~\ref{fig:strain_profiles_thin} shows similar results as Fig.~\ref{fig:strain_profiles_bulk} for the thin silicone layer. Here, an estimate of the average shear strain $\epsilon_{xz}$ in the layer is derived from the ratio of the measured in-plane displacement $u_x$ to the specimen thickness $h$.
%
As compared to the $u_x/a_0$ ratio for the bulk substrate, we observe that the shape of the average shear strain $u_x/h$ profiles is significantly altered when the normal load is increased (Fig.~\ref{fig:strain_profiles_thin}a)). Due to finite thickness, linear elasticity predicts an asymmetric shape skewed toward the leading edge regardless of the load. However, non-linear effects alter the scenario, and this is observed only at relatively high normal loads, when the contact area is sufficiently large for the confinement to affect the contact behavior. Conversely, at low loads, the $u_x/h$ distribution presents a skew toward the trailing edge, qualitatively recovering the bulk behavior. Interestingly, measured shear strains are always larger than predicted ones, with the ratio between the two being almost independent on the normal force. The same is not observed in the bulk case (predicted $u_x/a_0$ distributions are larger than the measured ones), confirming that the solid thickness is a game changer for the non-linear shear response.\\
In-plane strains present peak values at the trailing edge which are only slightly reduced compared to the bulk case (Fig.~\ref{fig:strain_profiles_thin}b)). This suggests that in both small strain and finite elasticity, strain concentration remains a local phenomenon, which is poorly affected by confinement, though layered solids should present a significantly stiffer response. 
%
\section{Conclusions}
The anisotropic shrinkage of contacts between silicon rubber and spherical glass probes under steady-state sliding was investigated for various levels of geometrical confinement. Regardless of the contact conditions, an asymmetry is systematically observed in the contact shape, which predominantly results from detachment of the silicone surface from the glass probe at the trailing edge of the contact. A comparison with linear contact mechanics calculations clearly indicates that the observed contact shrinkage is not ascribable to a friction-induced reduction of adhesion but to non-linear (finite-elasticity) phenomena triggered by frictional stresses, as confirmed by the measured values of in-plane strains (especially at contact edges) which lies well within the neo-Hookean range of the silicone elastomer and even slightly above.
Importantly, confinement enhances non-linear effects, especially regarding the anisotropic shrinkage of the contact area and, in general, the coupling between normal and tangential contact responses.
For instance, while for thick layers the stiction phase is associated with a lift-up of the glass probe under given normal load, \textit{i.e.} a reduction in the contact indentation, the opposite behavior (sink-in) is observed in confined conditions. 
Consistently, under imposed penetration conditions during stiction, the effective normal stiffness of the contact decreases as the confinement increases.
This non-linear coupling is usually overlooked in the experimental measurements of the contact area under shear; however, for testing arrangements unable to strictly control neither the normal load nor the displacement (\textit{e.g.}, double cantilever), the experiments are strongly affected by the apparatus stiffness. 
Overall, our results indicate that a deeper understanding of frictional polymeric contacts necessitates the inclusion of finite-elasticity effects, thus urging for the corresponding development of effective non-linear models and numerical methods. \textcolor{red}{Indeed, we are currently developing a non linear numerical framework to enable direct prediction of confined contacts. This will allow future work to connect our experimental findings on stiction and steady sliding with finite-strain theory in a fully quantitative manner}.
%
%
\section*{Acknowledgments}
This work has been partly supported by the European Union’s research and innovation programme Horizon Europe under the project (GC) “Handling with AI-Enhanced Robotic Technologies for Flexible Manufacturing (HARTU)“ - grant agreement No. 101092100, and by the European Union - NextGenerationEU through the Italian Ministry of University and Research under the programs (NM): PRIN2022 grant nr. 2022SJ8HTC - ELectroactive gripper For mIcro-object maNipulation (ELFIN); PRIN2022 PNRR grant nr. P2022MAZHX - TRibological modellIng for sustainaBle design Of induStrial friCtiOnal inteRfacEs
(TRIBOSCORE).
%

\end{document}